\documentclass[aps,prb,longbibliography,groupedaddress,twocolumn,10pt]{revtex4-2}
\usepackage{graphicx}
\usepackage{mathtools}
\usepackage{amsmath}
\usepackage{amssymb}
\usepackage{algorithm}
\usepackage{algpseudocode}
\usepackage{microtype}
\usepackage{bbm}
\usepackage{bm}
\usepackage{cancel}
\usepackage{xspace}
\usepackage{braket}
\usepackage{xcolor}
\usepackage{siunitx}
\usepackage{calrsfs}
\usepackage{comment}
\usepackage{placeins}
\usepackage[export]{adjustbox}
\usepackage{wrapfig}
\usepackage{bbold}
\usepackage{booktabs} 
\usepackage[colorlinks=true]{hyperref} 

\newcommand{\kel}[1]{\underline{#1}} 

\definecolor{hblue}{RGB}{0,80,255}
\definecolor{hred}{RGB}{255,80,0}

\begin{document}

\title{Steady-state study of the nonequilibrium properties of SrVO$_3$}

\author{Tommaso Maria Mazzocchi}
\email[]{mazzocchi@tugraz.at}
\author{Markus Aichhorn}
\author{Enrico Arrigoni}
\email[]{arrigoni@tugraz.at}
\affiliation{Institute of Theoretical and Computational Physics, Graz University of Technology, 8010 Graz, Austria}

\date{\today}


\begin{abstract}
We present the mixed-configuration approximation (MCA) based on the auxiliary master equation approach impurity solver to study multiorbital correlated systems under equilibrium and nonequilibrium conditions within dynamical mean-field theory (DMFT). We benchmark the method for bulk and layered SrVO$_3$ in equilibrium and apply it to a prototypical nonequilibrium geometry in which a voltage bias is applied perpendicular to the layer via reservoirs held at different chemical potentials. For bulk SrVO$_3$, MCA reproduces the metallic state at moderate interaction strengths, but it overestimates the weight of the lower band relative to quantum Monte Carlo (QMC) and fork tensor product state (FTPS) solvers. With respect to QMC and FTPS, MCA yields an earlier metal-to-insulator transition as the electron-electron interaction is increased. In layered SrVO$_3$ at equilibrium, MCA partially captures the orbital polarization in favor of the in-plane $xy$ orbital, although not as strong as in the DMFT-converged results obtained with QMC. Finally, under applied bias, we observe a pronounced redistribution of orbital occupations, demonstrating that the method captures bias-driven orbital charge transfer in realistic materials in nonequilibrium conditions.
\end{abstract}



\maketitle

\section{Introduction}\label{sec:intro}
Because of the interacting nature of their electrons, predicting the properties of strongly correlated systems represents a major challenge already in equilibrium. In recent years, interest in these materials has steadily grown. A contributing factor is represented by the discovery of the {\em resistive switching}, a phenomenon that consists in a metal-to-insulator transition as function of applied strains, studied from both theoretical~\cite{aron.12,am.we.12,mu.we.18,ma.ga.22,ha.ar.23} and experimental~\cite{ja.tr.15,ho.ad.17,su.ga.19} perspectives. Another aspect that boosted the field is the fact that, in contrast to its bulk form, SrVO$_3$ layers grown on a SrTiO$_3$ substrate become insulating when the film thickness is reduced to two layers~\cite{zh.wa.15}. This insulating behavior can persist up to three layers and has also been reported for a free-standing SrVO$_3$ monolayer~\cite{bh.as.16,ma.we.25}.
Being highly sensitive to external perturbations such as pressure, temperature, and electric field, the heterostructure SrVO$_3$ $+$ SrTiO$_3$ studied in Ref.~\cite{zh.wa.15} provides a suitable platform to investigate the Mott-Hubbard transition at a theoretical level. Moreover, it has been shown that applying a gate voltage can drive this system to switch between metallic and insulating states, which suggests a possible route toward transistor-like devices with a strong on-off contrast. For these reasons, strongly correlated materials have been considered as potential alternatives to conventional silicon-based electronic components.

A theoretical framework which has proven to be very effective for describing strongly correlated materials is the dynamical mean-field theory~\cite{ge.ko.92,ge.ko.96} (DMFT), especially when combined with {\em ab initio} methods such as the density functional theory~\cite{ho.ko.64,ko.sh.65} (DFT). The DMFT relies on the mapping of the correlated lattice problem onto an {\em impurity model} coupled to a reservoir of free electrons that has to be determined self-consistently. In this context, DMFT's only approximation consists in neglecting spatial correlation, while quantum fluctuations in time are treated exactly. 

The most computationally demanding step is solving the impurity problem, which is also the bottleneck of this approach. Several impurity solvers have been thus developed over the years. However, within DMFT, even relatively simple correlated materials often require many orbital degrees of freedom to be described correctly. This makes the problem more complex, since a multiband (or multiorbital) impurity solver is then needed, and it motivates the ongoing effort to develop more efficient solvers for many-orbital systems.

Among the available impurity solvers capable of treating multiorbital systems, quantum Monte Carlo~\cite{gu.mi.11,we.co.06} (QMC) methods are widely used. However, while they are exact (up to statistical errors), they can suffer from the fermionic sign problem in certain parameter regimes. In addition, QMC results are obtained on the imaginary-frequency axis and must be analytically continued to real frequencies, which is an ill-posed problem and, as such, introduces broadening of the spectral features, especially in the energy regions away from the Fermi level. Recent developments have shown that the so-called {\em inchworm} extension of the QMC algorithm can mitigate the sign problem~\cite{co.gu.15}. This scheme has then been employed to compute nonequilibrium properties of quantum systems within the Keldysh formalism~\cite{an.do.17}, also for multiorbital cases~\cite{er.bl.24}. However, to the best of our knowledge, even though inchworm QMC is very accurate, it cannot be used to reach arbitrarily long times, so addressing the steady-state dynamics is still challenging.

The numerical renormalization group~\cite{wils.75,bu.co.08} (NRG) is highly accurate at zero temperature and for energies close to the Fermi level, but its resolution becomes less reliable away from this low-energy window. The NRG has the advantage that it can also be extended to address nonequilibrium steady-state problems, see, e.g., Refs.~\cite{ande.08,han.25u}. However, recent work indicates that treating three correlated orbitals already pushes NRG close to its practical limits~\cite{st.mi.16,ho.zi.16}. 

Another important class of impurity solvers is based on exact diagonalization (ED)~\cite{ca.kr.94}. ED works directly on the real-frequency axis, but the Hilbert space grows exponentially with system size, which strongly limits the number of orbitals that can be treated. Compared to QMC and NRG, plain ED is often less accurate because the hybridization function can only be approximated with a finite number of bath sites, and good fits typically require many of them. However, it has been shown that, when embedded within the auxiliary master equation approach (AMEA) for open quantum systems~\cite{do.nu.14,we.lo.23}, ED becomes much more reliable.

Finally, matrix product states~\cite{scho.11,gr.ku.25} (MPS)-based techniques and the fork tensor product states~\cite{ba.zi.17} (FTPS) impurity solvers also represent a viable alternative. For nonequilibrium setups, influence-functional MPS methods can also reach long times and have recently been used to formulate steady-state impurity/DMFT schemes~\cite{na.th.25,so.li.25}. Even though these approaches can also work directly on the real-frequency axis, in practice the number of correlated orbitals that can be treated is often limited to a few, since the required bond dimension grows rapidly with increasing orbital count. For FTPS in particular, calculations with five correlated orbitals, as in the case of SrMnO$_3$ studied in Ref.~\cite{ba.tr.18}, appear to be close to the practical limit of what is currently feasible.

With no aim at comprehensiveness, we have shown that tackling systems of many interacting orbitals represents a challenge for all the methods discussed above. To address systems of many correlated orbitals in a rather computationally cheap way, in our previous work~\cite{ma.we.25} we developed an approximate method that we called mixed-configuration approximation (MCA). MCA assumes a good single-band impurity solver and can thus be applied to {\em any} algorithm, included MPS.

In Ref.~\cite{ma.we.25} we combined MCA with the AMEA impurity solver. As already mentioned, MCA can be coupled to any single-band impurity solver and, being based on AMEA, it can also deal with nonequilibrium setups. Its main approximation is to reduce the inter-orbital part of the Hubbard-Kanamori Hamiltonian to an effective single-particle term while still treating intra-orbital correlations exactly at the local level, as usual in DMFT. 
While the approach is based on multiple solutions of a correlated {\em single impurity} problem for a target orbital, it goes beyond a plain orbital mean-field approach by considering a weighted combination of all possible configurations of the other orbitals, see Sec.~\ref{sec:method}.

In this work, we extend the MCA-AMEA approach to multiorbital systems with an arbitrary number of orbitals. As it turns out, beyond two orbitals this procedure is highly nontrivial. We then explore applications to the case of bulk and layered SrVO$_3$ in and out of equilibrium. The paper is organized as follows. In Sec.~\ref{sec:model} we introduce the multiorbital impurity model. In Sec.~\ref{sec:method} we introduce the MCA and describe its generalization to an arbitrary number of orbitals. In Sec.~\ref{sec:results} we present and discuss our results. Finally, Sec.~\ref{sec:conclusions} contains our conclusions and an outlook for future development.

\section{The impurity model}\label{sec:model}

The Hamiltonian of the multiorbital impurity model in this work is given by
\begin{equation}\label{eq:H_tot}
\hat{H} = \hat{H}_{\text{imp}} + \hat{H}_{\text{bath}} + \hat{H}_{\text{i-b}},
\end{equation}
where $\hat{H}_{\text{imp}}$ is the Hubbard–Kanamori Hamiltonian in the density-density form,
\begin{align}\label{eq:H_imp}
\begin{split}
\hat{H}_{\text{imp}} & = \underbrace{\sum_{m\sigma} \varepsilon^{(0)}_{m\sigma} \hat{n}_{m\sigma}}_{\hat{H}_{0}} + \underbrace{\frac{U}{2} \sum_{m\sigma}\hat{n}_{m\sigma}\hat{n}_{m\overline{\sigma}}}_{\hat{H}_{\text{\tiny intra}}} \\
& + \underbrace{\frac{U^{\prime}}{2} \sum_{\substack{m \neq n}} \sum_{\sigma} \hat{n}_{m\sigma} \hat{n}_{n\overline{\sigma}} + \frac{U^{\prime\prime}}{2} \sum_{\substack{m \neq n}} \sum_{\sigma} \hat{n}_{m\sigma} \hat{n}_{n\sigma}}_{\hat{H}_{\text{\tiny inter}}}.
\end{split}
\end{align}
In the Hamiltonian in Eq.~\eqref{eq:H_imp}, we neglect spin-flip and pair-hopping terms and choose an orbital basis in which the single-particle Hamiltonian is diagonal, so that inter-orbital hopping is absent. This approximation is expected to be reliable when these contributions are small, e.g., at low filling or when the DFT-projected Hamiltonian is (approximately) diagonal in the chosen basis, as in the present work.

In Eq.~\eqref{eq:H_imp}, $\varepsilon^{(0)}_{m\sigma}$ is the on-site energy of orbital $m$, $\hat{n}_{m\sigma} \left( \equiv \hat{c}^{\dagger}_{m\sigma} \hat{c}_{m\sigma}\right)$ the corresponding density operator and $\overline{\sigma}\equiv -\sigma$. In this work we choose $U^{\prime} = U - 2J$ and $U^{\prime\prime} = U^{\prime} - J$, where $J$ is the Hund's coupling.

The impurity site is connected to a fermionic reservoir described by the Hamiltonian
\begin{equation}
\hat{H}_{\text{bath}} = \sum_{mq\sigma} \varepsilon_{mq} \hat{f}^{\dagger}_{mq\sigma} \hat{f}_{mq\sigma},
\end{equation}
where $q$ labels the (infinite) energy levels of the reservoir and \( \hat{f}^{\dagger}_{mq\sigma} \) (\( \hat{f}_{mq\sigma} \)) the creation (annihilation) operator of an electron with spin \( \sigma \) with energy \(\varepsilon_{mq}\).
In this work, we restrict to diagonal coupling between the orbitals on the impurity and the electronic levels in the reservoir, i.e.
\begin{equation}
\hat{H}_{\text{i-b}} = \sum_{mq\sigma}  V_{mq} \left( \hat{c}^{\dagger}_{mq\sigma} \hat{f}_{mq\sigma} + \text{H.c.} \right).
\end{equation}
The reservoir can be completely characterized by the so-called {\em hybridization function}~\cite{do.nu.14,we.lo.23,ma.we.25}, which, for the systems addressed in this work, is diagonal in orbital and spin sectors.

\section{Methods}\label{sec:method}

In this section, we introduce a mixed-configuration approximation (MCA) for an impurity with \(N\) correlated orbitals. In MCA, the orbital space is partitioned into a {\em target} orbital, the Green's function (GF) of which is computed explicitly, and the remaining {\em configuration} orbitals, which are treated through their possible occupation probabilities.

The impurity GF of the target orbital is obtained as a weighted sum over the GFs of a set of independent single-impurity problems, each corresponding to a fixed configuration of the other orbitals~\footnote{A configuration is defined as the collection of fixed occupation states of the configuration orbitals.}. The weights in this sum are the joint probabilities of finding the configuration orbitals in the corresponding configuration. Consequently, computing the GF for a given target orbital requires solving one impurity problem for each inequivalent configuration; this procedure has to be repeated for all the correlated orbitals in the impurity.

To solve these independent single-impurity problems, we employ the auxiliary master equation approach~\cite{do.nu.14,we.lo.23} (AMEA) impurity solver~\footnote{As already argued in Ref.~\cite{ma.we.25}, any single-band impurity solver can be extended according to the prescription of the MCA scheme.}, an efficient steady-state method which has proven accurate in several nonequilibrium settings, see, e.g., \cite{so.do.18,ma.ga.22,ga.ma.22,we.lo.23,ma.we.23,ga.we.24,ku.er.24} for details.

We now present the basic assumptions of the MCA. Without loss of generality, we choose orbital \(m=1\) as the target orbital; the same reasoning applies to any other choice of \(m\). We denote the set of impurity orbitals by
\(\mathcal{O}=\{ 1, \dots, N \}\), and the set of configuration orbitals by \(\bar m \equiv \mathcal{O}\setminus\{m\}\).

Each of the orbitals can be in one of the four states \(\alpha\in\mathcal{S}\equiv\{\mathrm{e},\uparrow,\downarrow,\mathrm{d}\}\), where \(\mathrm{e}\) denotes the {\em empty} and \(\mathrm{d}\) the {\em doubly occupied} state. A configuration is specified by the tuple of the states of the configuration orbitals,
\begin{equation}\label{eq:conf_short}
C \equiv (l_{\alpha_l})_{l \in \bar m},
\end{equation}
where $\alpha_{l} \in \mathcal{S}$. For instance, by choosing the target orbital \(m=1\) one has \(C=(2_{\alpha_2}, 3_{\alpha_3}, \dots, N_{\alpha_N})\).

In MCA, we evaluate the GF of the target orbital for each configuration $C$ of the configuration orbitals. To this end, we replace the density operators of the configuration orbitals, \(\hat n_{l\sigma}\) with \(l\in\bar m\), by numbers \(n_{l\sigma}(C)\in\{0,1\}\) determined by the configuration \(C\). 
To define them, we introduce the spin set associated with a state,
\begin{equation}\label{eq:occupancy_set}
S(\alpha_{l})=
\begin{cases}
\emptyset, & \alpha_{l}=\mathrm{e}, \\[4pt]
\{\uparrow\}, & \alpha_{l}=\uparrow, \\[4pt]
\{\downarrow\}, & \alpha_{l}=\downarrow, \\[4pt]
\{\uparrow,\downarrow\}, & \alpha_{l}=\mathrm{d},
\end{cases}
\end{equation}
and the corresponding indicator (characteristic) function
\begin{equation}\label{eq:indicator}
n_{l\sigma}(C)\equiv \mathbf{1}_{\sigma\in S(\alpha_l)} \equiv
\begin{cases}
1, & \text{if } \sigma \in S(\alpha_l),\\
0, & \text{if } \sigma \notin S(\alpha_l).
\end{cases}
\end{equation}

With this replacement, the inter-orbital part of the Hamiltonian~\eqref{eq:H_imp} becomes a single-particle operator, dependent on the configuration \(C\). 
This produces a contribution to the on-site energy of the target orbital \(m\) which becomes
\begin{equation}\label{eq:generate_eps_correction}
\varepsilon_{m\sigma}(C)
=
\varepsilon^{(0)}_{m\sigma}
+
\sum_{l\in\bar m}\Big[ U^{\prime}\, n_{l\bar\sigma}(C) + U^{\prime\prime}\, n_{l\sigma}(C)\Big],
\end{equation}
with $n_{l\sigma}(C)$ the occupation number of orbital \(l\) in state \(\sigma\) for the configuration \(C\).
We next determine the weights with which the configuration-dependent GFs for the target orbital \(m\) determined in this way contribute to its total GF. The {\em conditional} probabilities that the target orbital is found in a particular state \(\alpha_m\in\mathcal{S}\) for a given configuration \(C\) are calculated from the occupations obtained from the solution of the corresponding single-impurity problem with the energies~\eqref{eq:generate_eps_correction} as
\begin{align}\label{eq:conditional_probs_2orb}
\begin{split}
\bar P(m_{\mathrm{d}}\mid C) &= \big\langle \hat n_{m\uparrow}\hat n_{m\downarrow}\big\rangle_{C}, \\
\bar P(m_{\uparrow}\mid C) &= \big\langle \hat n_{m\uparrow}\big\rangle_{C} - \bar P(m_{\mathrm{d}}\mid C), \\
\bar P(m_{\downarrow}\mid C) &= \big\langle \hat n_{m\downarrow}\big\rangle_{C} - \bar P(m_{\mathrm{d}}\mid C), \\
\bar P(m_{\mathrm{e}}\mid C) &= 1 - \sum_{\alpha \in \mathcal{S} \setminus \mathrm{e}} \bar P(m_{\alpha}\mid C),
\end{split}
\end{align}
where \(\langle\cdots\rangle_{C}\) denotes the expectation value of one operator calculated from the impurity solver for the fixed configuration \(C\)~\footnote{Our notation is such that $\bar P$ denotes conditional probability.}. By construction, the conditional probabilities are normalized,
\begin{equation}\label{eq:column_stochastic_matrix}
\sum_{\alpha\in\mathcal{S}} \bar P(m_{\alpha}\mid C)=1.
\end{equation}
The {\em marginal} (i.e. total) probabilities for the occupation of orbital \(m\) are obtained from
\begin{equation}\label{eq:total_probs_2orb}
P(m_{\alpha}) = \sum_{C} \bar P(m_{\alpha}\mid C)\, P_{\text{\tiny J}}(C),
\end{equation}
where \(P_{\text{\tiny J}}(C)\) denotes the \emph{joint} probability of finding the other orbitals in the configuration \(C\) defined as in Eq.~\eqref{eq:conf_short}.
These joint probabilities are the weights required to compute the (orbital) impurity GFs as
\begin{equation}\label{eq:imp_GFs_2orb}
\kel{G}_{m\sigma}(\omega) = \sum_{C} P_{\text{\tiny J}}(C)\, \kel{G}_{m\sigma}(\omega;C).
\end{equation}
The next section is devoted to the derivation of the joint probabilities \(P_{\text{\tiny J}}(C)\).

While this scheme can be applied in equilibrium, in our nonequilibrium steady state case the configuration-dependent impurity GFs on the right-hand side of Eq.~\eqref{eq:imp_GFs_2orb} satisfy the Keldysh structure~\cite{ma.ga.22,ma.we.23}
\begin{equation}\label{eq:keldysh_structure}
\kel{G}=
\begin{pmatrix}
G^{\text{R}} & G^{\text{K}} \\
0 & G^{\text{A}}
\end{pmatrix},
\end{equation}
where \(G^{\text{R}}\), \(G^{\text{K}}\), and \(G^{\text{A}}=\big[G^{\text{R}}\big]^{\dagger}\) are the \emph{retarded}, \emph{Keldysh}, and \emph{advanced} components, respectively.

\subsection{The algorithm for arbitrary $N$}\label{sec:iterative_probs}

In this section we will show the procedure to determine the joint probabilities \(P_{\text{\tiny J}}(C)\) once orbital \(m\) is targeted.
For a fixed target \(m=1\), we denote a configuration of the remaining \( \left( N - 1 \right) \) orbitals by the tuple~\footnote{The configuration tuple resulting by singling out only one target orbital lists all the remaining orbitals with their respective states. We point out that given the correction to the onsite energy in Eq.~\eqref{eq:generate_eps_correction}, the order of the orbitals in the configuration does not matter.}
\begin{equation}
C \equiv (2_{\alpha_2}, 3_{\alpha_3}, \dots, N_{\alpha_{N}}).
\end{equation}
The marginal probability of orbital \(1\) (in state \( \alpha_{1} \in \mathcal{S} \)) is obtained as
\begin{align}\label{eq:orb1_gen_procedure_Ps}
\begin{split}
P(1_{\alpha_1}) & = \sum_{C} \bar P(1_{\alpha_1}\mid C) P_{\text{\tiny J}}(C), \\
& = \sum_{\alpha_{2}\alpha_{3}\dots \alpha_{N}} \bar P(1_{\alpha_1}\mid 2_{\alpha_2}, 3_{\alpha_3}, \dots, N_{\alpha_{N}}) \\
& \times P_{\text{\tiny J}}(2_{\alpha_2}, 3_{\alpha_3}, \dots, N_{\alpha_{N}})
\end{split}
\end{align}
where \( \bar P(1_{\alpha_1}\mid C)\) is obtained from the corresponding single-impurity problem with target orbital \(1\) and configuration \(C\), and \(P_{\text{\tiny J}}(C)\) is the (unknown) joint probability of the configuration orbitals. Notice that the sum schematically denoted in Eq.~\eqref{eq:orb1_gen_procedure_Ps} must be understood as \( \sum_{C} \equiv \sum_{(l_{\alpha_l})_{l \in \bar 1}} \), where \( \bar 1 = \{ 2, 3, \dots, N \}\) and \( \alpha_{l} \in \mathcal{S} \), according to the definition in~\eqref{eq:conf_short}, as in the second equality in~\eqref{eq:orb1_gen_procedure_Ps}.

Using the marginalization rule,
\begin{equation}\label{eq:PJ_marginalization}
P_{\text{\tiny J}}(C) = \sum_{\alpha_1'\in\mathcal{S}} \bar P(C \mid 1_{\alpha_1'}) P(1_{\alpha_1'}),
\end{equation}
Eq.~\eqref{eq:orb1_gen_procedure_Ps} can be recast as the fixed-point equation~\footnote{Notice that solving Eq.~\eqref{eq:P_marginal_1} is equivalent to finding the eigenvector corresponding to the eigenvalue 1 of the matrix~\eqref{eq:full_cp_matrix}. The existence of one such eigenvalue is guaranteed by the {\em column stochastic} nature of the conditional probability matrices, see Ref.~\cite{ma.we.25}. Resorting to the fixed-point method, i.e. iterating Eq.~\eqref{eq:P_marginal_1} until the probability {\em vector} does not change any longer, constitutes a more stable and versatile option than solving the eigenvalue problem, especially in (quasi-) degenerate systems.}
\begin{equation}\label{eq:P_marginal_1}
P(1_{\alpha_1})
=
\sum_{\alpha_1'\in\mathcal{S}}
\left[M\right]_{\alpha_1,\, \alpha_1'}\,
P(1_{\alpha_1'}),
\end{equation}
with
\begin{equation}\label{eq:full_cp_matrix}
\left[M\right]_{\alpha_1,\, \alpha_1'}
\equiv
\sum_{C}
\bar P (1_{\alpha_1}\mid C)\,
\bar P (C\mid 1_{\alpha_1'}).
\end{equation}
We notice that, for \(N=2\), the configuration \(C\) contains only one orbital, i.e. $C \equiv 2_{\alpha_2}$. In this case, \( \bar P(C \mid 1_{\alpha_1'}) \equiv \bar P(2_{\alpha_2} \mid 1_{\alpha_1'})\) can then be obtained directly from the impurity solver, reducing the procedure to the one described in our previous work~\cite{ma.we.25}. The algorithm to calculate \( \bar P(C \mid 1_{\alpha_1'})\) for \(N>2\) will be presented below.

It is worth mentioning that when $C$ contains two (or more) orbitals, $\bar P(C \mid 1_{\alpha_1'})$ cannot be calculated from a single impurity solver, as it would require the exact treatment of an impurity consisting of two (or more) correlated orbitals.

In addition, we point out that the occupations computed from the GFs are in one-to-one correspondence with those obtained from the probabilities $P(i_{\alpha_i})$. The latter can therefore be interpreted as analogues of the density-matrix elements associated with the local Hilbert space of a single orbital.

\paragraph{Evaluation of \( \bar P(C\mid 1_{\alpha_1'})\).} 

Here, we illustrate the procedure explicitly for the case of $N=3$ orbitals. In this case, the configuration of the other $N-1$ orbitals reads $C=(2_{\alpha_{2}}, 3_{\alpha_{3}})$, so that the conditional probability we look for can be recast as 
\begin{equation}\label{eq:P_tilde}
\bar P(C\mid 1_{\alpha_1'}) = \bar P(2_{\alpha_{2}}, 3_{\alpha_{3}} \mid 1_{\alpha_1'}).
\end{equation}
$\bar P(C\mid 1_{\alpha_1'})$ can be found with the help of the chain rule for conditional probabilities, i.e.
\begin{equation}\label{eq:P_tilde_chain_rule}
\bar P(2_{\alpha_{2}}, 3_{\alpha_{3}} \mid 1_{\alpha_1'}) = \bar P(2_{\alpha_{2}} \mid 3_{\alpha_{3}}, 1_{\alpha_1'}) \bar P(3_{\alpha_{3}} \mid 1_{\alpha_1'}).
\end{equation}
It is worth mentioning that $\bar P(2_{\alpha_{2}} \mid 3_{\alpha_{3}}, 1_{\alpha_1'})$ are determined according to the procedure described in Eq.~\eqref{eq:conditional_probs_2orb}, i.e. by {\em targeting} orbital 2 and solving as many (single) impurity solvers as there are independent configurations of the orbitals 1 and 3. 

At this point, we need to determine the conditional probabilities $P(3_{\alpha_{3}} \mid 1_{\alpha_1'})$. These probabilities contains only 2 orbitals (1 and 3), which suggests they can be obtained by a fixed point iteration summing over the states of orbital 2
\begin{align}\label{eq:tensor_contraction}
\begin{split}
\bar P(3_{\alpha_{3}} \mid 1_{\alpha_1'}) & = \sum_{\alpha_2'\alpha_3'} \bar P(3_{\alpha_{3}} \mid 2_{\alpha_2'} 1_{\alpha_1'}) \bar P(2_{\alpha_2'} \mid 3_{\alpha_3'} 1_{\alpha_1'}) \\
& \times \bar P(3_{\alpha_3'} \mid 1_{\alpha_1'}).
\end{split}
\end{align}
For each (fixed) value of the state of orbital 1 ($\alpha_1'$), Eq.~\eqref{eq:tensor_contraction} can be solved by a fixed point iteration. In alternative, one can converge the whole two-index tensor (matrix) $\bar P(3_{\alpha_{3}} \mid 1_{\alpha_1'})$ iteratively. 

Once $\bar P(3_{\alpha_{3}} \mid 1_{\alpha_1'})$ has been obtained using~\eqref{eq:tensor_contraction}, it can be used to compute the conditional probabilities in Eq.~\eqref{eq:P_tilde_chain_rule}, which in turn are used to calculate the matrix~\eqref{eq:full_cp_matrix}, to find the marginal probabilities $P(1_{\alpha_1})$ in Eq.~\eqref{eq:P_marginal_1} that describe the occupation of orbital 1.

Once $P(1_{\alpha_1})$ are known, the joint probability of the two remaining orbitals can be computed as described in~\eqref{eq:PJ_marginalization}, which in turn will be used to compute the orbital impurity GF as in Eq.~\eqref{eq:imp_GFs_2orb}.

The procedure described here needs to be repeated for all the other orbitals in the impurity, i.e. by targeting orbitals 2 and 3. By increasing the number of orbitals one has to deal with conditional probabilities with more orbitals, which increases the complexity of the algorithm. For instance, in the case of $N=4$ orbitals, one has $\bar P(C\mid 1_{\alpha'}) = \bar P(2_{\alpha_2}, 3_{\alpha_3}, 4_{\alpha_4} \mid 1_{\alpha_1'})$.

Using the chain rule, these probabilities can be recast as
\begin{align}\label{eq:P_tilde_chain_rule_4orbs}
\begin{split}
\bar P(2_{\alpha_2}, 3_{\alpha_3}, 4_{\alpha_4} \mid 1_{\alpha_1'}) & = \bar P(2_{\alpha_2} \mid 3_{\alpha_3}, 4_{\alpha_4}, 1_{\alpha_1'}) \\
& \times \bar P(3_{\alpha_3} \mid 4_{\alpha_4}, 1_{\alpha_1'}) \bar P(4_{\alpha_4} \mid 1_{\alpha_1'}).
\end{split}
\end{align}
Once more, the quantities $\bar P(3_{\alpha_3} \mid 4_{\alpha_4}, 1_{\alpha_1'})$ and $\bar P(4_{\alpha_4} \mid 1_{\alpha_1'})$ in Eq.~\eqref{eq:P_tilde_chain_rule_4orbs} are determined by fixed point iteration from the equations
\begin{align}
\begin{split}
\bar P(3_{\alpha_3} \mid 4_{\alpha_4}, 1_{\alpha_1'}) & = \sum_{\alpha_2'\alpha_3'} \bar P(3_{\alpha_3} \mid 2_{\alpha_2'}, 4_{\alpha_4}, 1_{\alpha_1'}) \\
&  \bar P(2_{\alpha_2'} \mid 3_{\alpha_3'}, 4_{\alpha_4}, 1_{\alpha_1'}) \bar P(3_{\alpha_3'} \mid 4_{\alpha_4}, 1_{\alpha_1'}),
\end{split}
\end{align}
and
\begin{align}
\begin{split}
\bar P(4_{\alpha_4} \mid 1_{\alpha_1'}) & = \sum_{\alpha_3'} \bar P(4_{\alpha_4} \mid 3_{\alpha_3'}, 1_{\alpha_1'}) \\
& \bar P(3_{\alpha_3'} \mid 4_{\alpha_4}, 1_{\alpha_1'}) \bar P(4_{\alpha_4} \mid 1_{\alpha_1'}).
\end{split}
\end{align}

The computational cost of MCA-AMEA scales exponentially in $N-1$ (without symmetries). In particular, our approximation leaves the Hilbert space unchanged, since it treats only a single orbital at a time. By contrast, in a full multiorbital AMEA impurity solver calculation, the size of the Hilbert space grows exponentially with $2(N_{\text{B}}+1)N$, rendering calculations with $N_{\text{B}}=6$ or $8$ bath sites practically infeasible.

Once the multiorbital nonequlibrium steady-state Keldysh GF is computed, the self-consistent DMFT procedure is standard and will not be discussed here. For details about DMFT, we point the reader to Refs.~\cite{ge.ko.92,ge.ko.96}. For the nonequilibrium extension of DMFT, the work~\cite{ao.ts.14} provides a comprehensive description, while in Refs.~\cite{ma.ga.22,ga.ma.22,ma.we.23,ma.we.25} more applications of DMFT to nonequilibrium steady-state problems can be found. Notice that the approach works in the symmetric case only in which no orbital-mixing terms arise from the DMFT self-consistency.

\section{Results}\label{sec:results}

In this section we present and discuss the DMFT-converged results obtained with our MCA-AMEA solver both in equilibrium and nonequilibrium setups. 
We start with the equilibrium case of bulk SrVO$_3$. Later we move on to the investigation of the layered structure already studied in Ref.~\cite{bh.as.16} and, limited to a two-orbital setup, in Ref.~\cite{ma.we.25}, both in and out of equilibrium conditions.

For the equilibrium benchmarks, we compare the results obtained within MCA-AMEA against those obtained with QMC. As QMC solver, we employ continuous time quantum Monte Carlo in the hybridisation expansion~\cite{we.co.06} as implemented in the TRIQS/CTHYB package~\cite{pa.fe.15,triqscthyb}. In addition, the DFT Hamiltonians of both the bulk and monolayer SrVO$_3$ have been constructed using the Wien2k (version 14.2)~\cite{Blaha_2020_PAPER}, TRIQS/DFTTools (version 3.3)~\cite{pa.fe.15,Aichhorn2016}, and Wannier90 (version 3.0)~\cite{Pizzi2020} packages.

\subsection{Bulk SrVO$_3$ at equilibrium}\label{sec:bulk_strontium_vanadate}

We start this section with bulk SrVO$3$ at equilibrium. This is arguably the most challenging test for MCA, due to the full degeneracy of the orbitals in the $t_{2g}$ manifold.
As already discussed in our previous work~\cite{ma.we.25}, MCA tends to shift the spectral weight away from the Fermi level when compared to other solvers such as QMC in systems involving (almost) degenerate orbitals. However, the overall accuracy of our method improves when the orbital degeneracy is (partly) lifted, as in the case of the layered SrVO$_3$, which will be discussed in Sec.~\ref{sec:layered_strontium_vanadate}.

\begin{figure}[h]
\includegraphics[width=\linewidth]{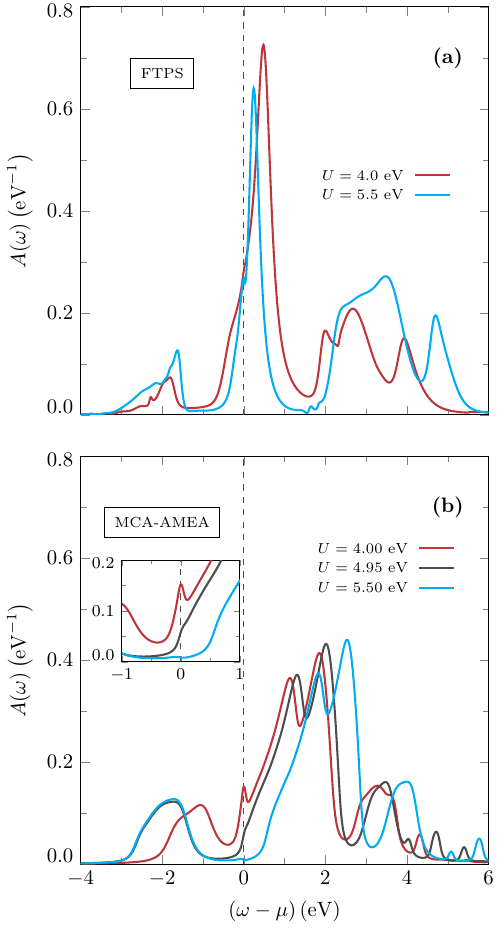}
\caption{Equilibrium spectra of bulk SrVO$_3$ obtained for selected values of $U$ and $J=0.7 \, \mathrm{eV}$. Results obtained within the (a) FTPS (figure adapted from \href{https://doi.org/10.1103/PhysRevX.7.031013}{Phys. Rev. X \textbf{7}, 031013 (2017)}) and (b) MCA-AMEA. The inset in (b) magnifies the region around the Fermi level, while the dashed vertical lines denote the position of the chemical potential.}
\label{fig:fig1}
\end{figure}

In Fig.~\ref{fig:fig1} we compare the equilibrium spectra of SrVO$_3$ obtained with the FTPS solver~\cite{ba.zi.17} and our MCA-AMEA approach for several values of the Hubbard interaction $U$. In its bulk form, SrVO$_3$ is a \( d^{1} \) metal, with one electron distributed over the three \( t_{2g} \) orbitals. The FTPS results, see Fig.~\ref{fig:fig1}(a), show that bulk SrVO$_3$ remains metallic from intermediate up to large interaction strengths, a result which is in agreement with the literature~\cite{lieb.03,se.fu.04,pa.bi.04,ne.ke.05,ne.he.06,no.ka.12,zh.wa.15}. In contrast, the spectra obtained with MCA-AMEA display a clear tendency toward becoming an insulator already at smaller values of $U$, see Fig.~\ref{fig:fig1}(b).

In particular, for $U \geq 5 \ \mathrm{eV}$ a gap opens at the Fermi level (the inset in Fig.~\ref{fig:fig1}(b) highlights the low-energy region). The FTPS spectra show a mild tendency toward sharper band edges as $U$ increases but the density of states (DOS) preserves a consistent spectral weight at the Fermi level even at large values of $U$. On the other hand, within MCA-AMEA the DOS does not develop any sharp feature, as visible in Fig.~\ref{fig:fig1}(b).

For $U=4 \, \mathrm{eV}$, the lower Hubbard band (LHB) obtained with MCA-AMEA is shifted to slightly higher energies compared to FTPS, while still featuring a {\em shoulder} close to the Fermi level, which is however more pronounced than in the FTPS result. 

As we mentioned above, given the limitations of our method in dealing with systems of degenerate orbitals, the difficulty of MCA-AMEA to retain a metallic solution for bulk SrVO$_3$ at large $U$ is expected. This behavior is evident in Fig.~\ref{fig:fig1}(b): already at $U=4.95 \, \mathrm{eV}$, the LHB carries too much spectral weight, thus forcing the Fermi level to readjust and lie in the gapped region. This property becomes stronger as $U$ is increased further. In the next section we will present the results for a case in which the orbital degeneracy is partially lifted and MCA performs better.

\subsection{Layered SrVO$_3$ at equilibrium}\label{sec:layered_strontium_vanadate}

\begin{figure*}[t]
\includegraphics[width=\linewidth]{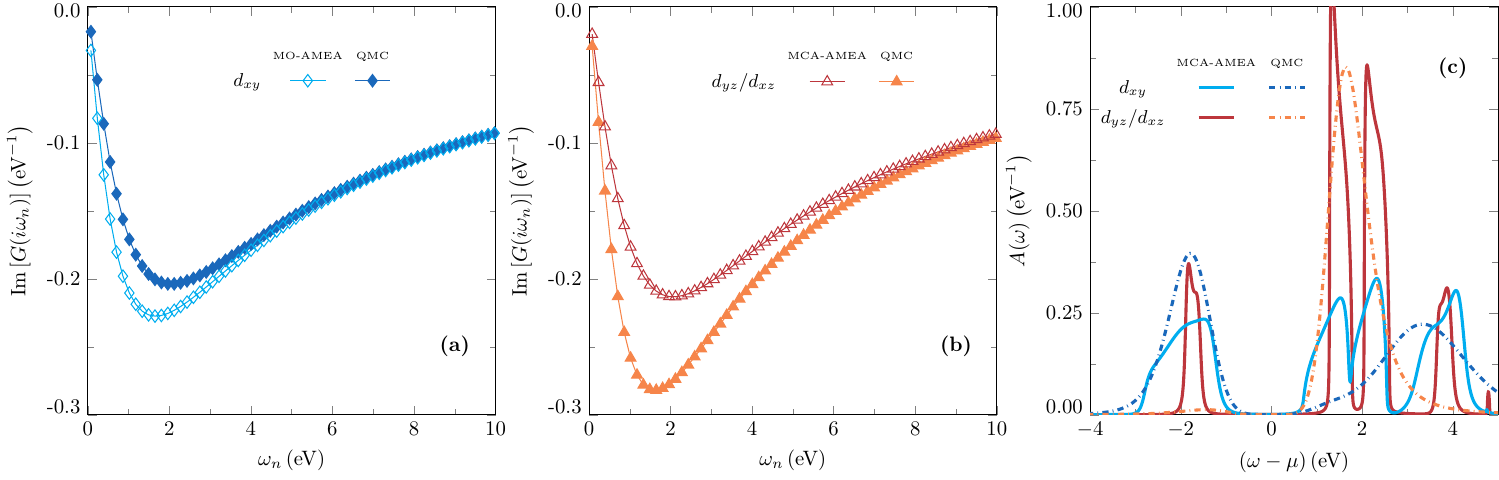}
\caption{DMFT equilibrium results for a free-standing monolayer of SrVO$_3$ at $1/6$ filling. (a) Imaginary part of the Matsubara GFs per spin for orbital $d_{xy}$ obtained with QMC and MCA-AMEA (The AMEA calculation is carried out in real frequencies, so the Matsubara data are obtained by the usual Laplace-type transform.). (b) Same for orbitals $d_{yz}$ and $d_{zx}$. (c) Orbital spectra obtained with QMC and MCA-AMEA. MCA-AMEA yields occupations $n_{d_{xy}} = 0.307$ and $n_{d_{yz}/d_{zx}} = 0.117$, hinting at charge polarization. By contrast, QMC predicts an almost fully polarized state with $n_{d_{xy}} = 0.462$ and $n_{d_{yz}/d_{zx}} = 0.019$ at numerically exact sixth-filling. Here $U = 5.5\, \mathrm{eV}$, $J = 0.75\, \mathrm{eV}$ and $T = 0.025\, \mathrm{eV}$.}
\label{fig:fig2}
\end{figure*}

When grown in thin films, SrVO$_3$ exhibits an insulating phase~\cite{zh.wa.15,bh.as.16} that can survive up to three-layer thick structures~\cite{zh.wa.15}. The reason is the partial lift of the degeneracy in the \( t_{2g} \) manifold due to the establishment of a {\em crystal field splitting}. This, in turn, results into a charge-polarized state where the in-plane \( xy \) orbital is (almost) completely full and the other (degenerate) \( yz/zx \) orbitals (almost) completely empty.

In Fig.~\ref{fig:fig2}(a)-(b) we compare the DMFT-converged imaginary parts of the Matsubara GFs in the \( t_{2g} \) sector for layered SrVO$_3$ at equilibrium, obtained with QMC and MCA-AMEA. In both approaches, $\mathrm{Im}\,G(i\omega_n)$ extrapolates to zero as $\omega_n \to 0$, consistent with an insulating solution. The best agreement between the two solvers is found for the in-plane \( d_{xy} \) orbital, whereas larger deviations appear for the remaining two (degenerate) orbitals.

The corresponding real-frequency spectra are shown in Fig.~\ref{fig:fig2}(c). MCA-AMEA reproduces the size of the insulating gap rather well, and the band positions are also in close agreement with QMC. While MCA-AMEA underestimates the magnitude of the orbital polarization, it still captures the trend toward an enhanced occupation of the $d_{xy}$ orbital (see the caption of Fig.~\ref{fig:fig2} for the numerical values).

At this stage, it is not clear whether the reduced polarization obtained with MCA-AMEA originates mainly from the DMFT self-consistency cycle, e.g., through a gradual drift of the hybridization function, or whether it reflects the intrinsic limitations of the approximation. Compared to the two-orbital two-dimensional model studied in Ref.~\cite{ma.we.25}, the agreement with QMC indeed deteriorates when a third orbital is included, suggesting that the approximation becomes less accurate as the number of orbitals increases, most likely for the limitation of resolving the degeneracy, as remarked in Sec.~\ref{sec:bulk_strontium_vanadate}.

\begin{figure*}[t]
\includegraphics[width=\linewidth]{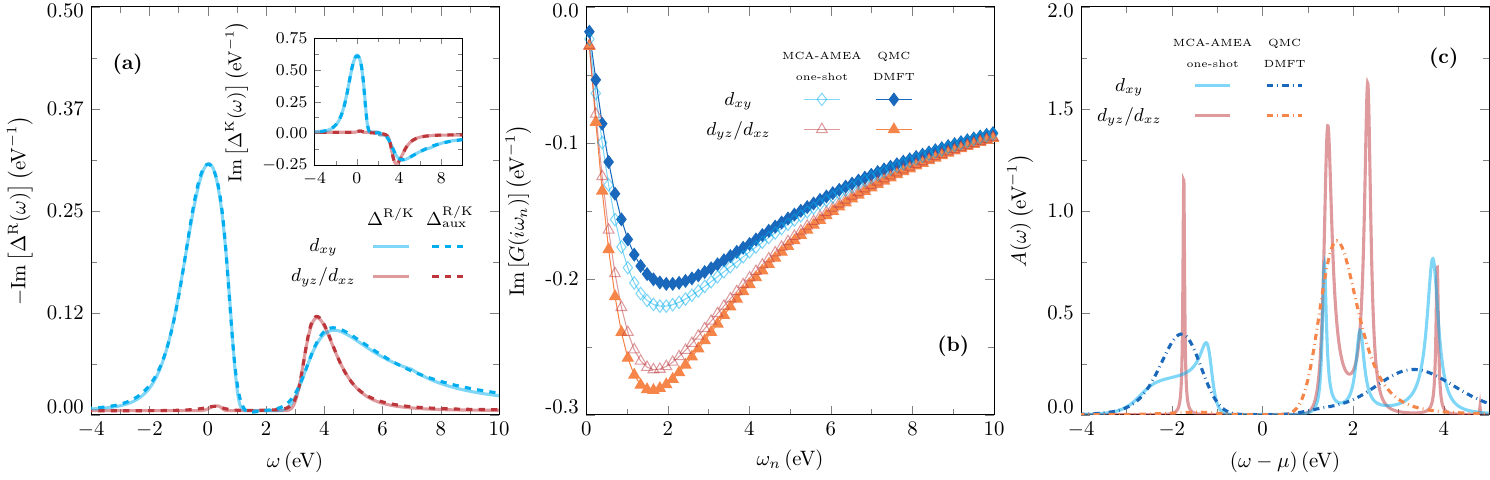}
\caption{Comparison of the one‐shot MCA‐AMEA results against fully self‐consistent DMFT obtained with QMC. (a) Orbital-resolved {\em retarded} hybridization function (solid) from the converged DMFT run within QMC used to initialize the one‐shot MCA‐AMEA calculation. The dashed line shows the reconstructed {\em auxiliary} hybridization function \( \Delta^{\text{R}}_{\text{aux}}\) necessary for the MCA-AMEA impurity solver. The inset shows the corresponding Keldysh components. (b) Imaginary part of the orbital Matsubara GFs. (c) Corresponding spectra to (b). MCA‐AMEA yields orbital occupations (per spin) \( n_{d_{xy}} = 0.372 \) and \( n_{d_{yz}/d_{zx}} = 0.078 \), hinting at a tendency toward a charge polarization in favor of orbital \(d_{xy}\). By contrast, the occupations in the converged QMC are \( n_{d_{xy}} = 0.462 \) and \( n_{d_{yz}/d_{zx}} = 0.019 \), with numerically exact $1/6$ filling. Temperature is set to \(T = 0.025\ \mathrm{eV}\), and the chemical potential is set to the last value from the DMFT‐converged result obtained with QMC, \(\mu= 2.09\ \mathrm{eV}.\)}
\label{fig:fig3}
\end{figure*}

To disentangle these effects, we also perform a one-shot impurity calculation using the QMC-converged hybridization function shown in Fig.~\ref{fig:fig3}(a) as input. In this case, MCA-AMEA shows a noticeably improved agreement with QMC on both the Matsubara axis and in real frequency for all orbitals, see Fig.~\ref{fig:fig3}(b) and (c). Consistent with this improvement, MCA-AMEA also yields a stronger tendency toward orbital polarization in favor of $d_{xy}$, although it still does not reach the fully polarized values obtained with QMC (see the caption of Fig.~\ref{fig:fig3}). 

We conclude that the differences between the DMFT-converged results obtained with MCA-AMEA and QMC are to be attributed to the combination of the approximate nature of MCA and the gradual drift of the hybridization that occurs during the self-consistent cycle that ultimately allows the DMFT to reach a different stationary point.

\subsection{Layered SrVO$_3$ in nonequilibrium conditions}\label{sec:neq_strontium_vanadate}

Here we consider a prototypical nonequilibrium setup in which the layered structure introduced in Sec.~\ref{sec:layered_strontium_vanadate} is sandwiched between two metallic contacts held at different chemical potentials. The left and right potentials are shifted symmetrically with respect to the equilibrium chemical potential of the configuration in Fig.~\ref{fig:fig3}, generating a voltage bias
$\Phi \equiv \mu_{\mathrm{l}} - \mu_{\mathrm{r}}$.
This bias drives a steady-state current $J_{\mathrm{l}\to\mathrm{r}}$ perpendicular to the layers~\cite{ma.we.25}; its explicit expression can be found in Refs.~\cite{ga.ma.22,ma.we.25}.

\begin{figure}[h]
\includegraphics[width=\linewidth]{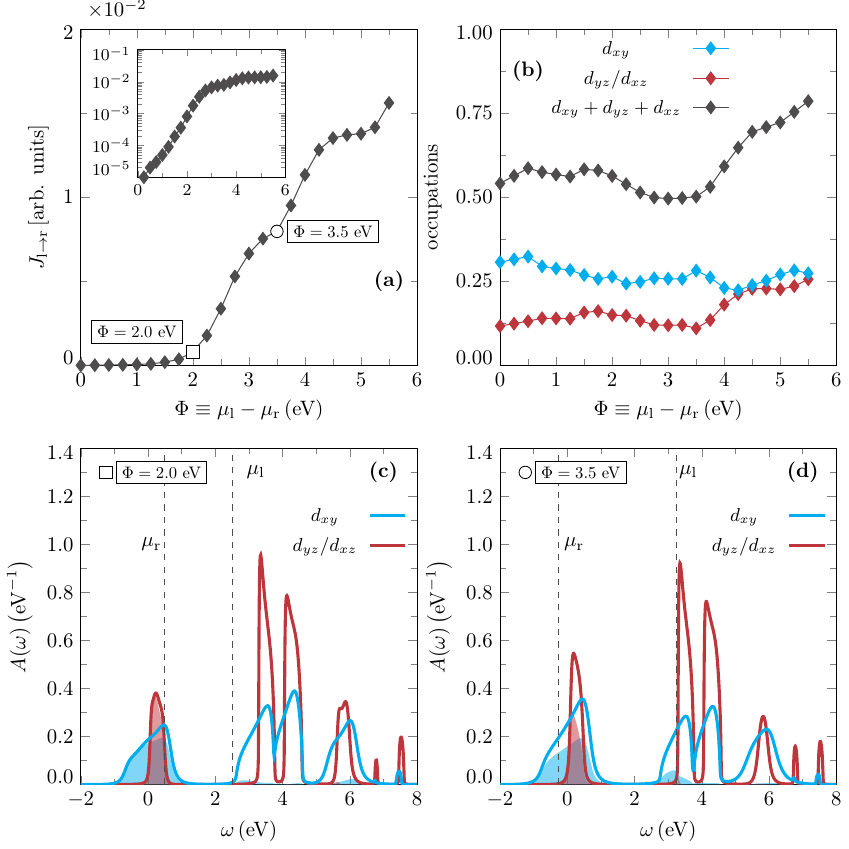}
\caption{Nonequilibrium DMFT results obtained with the MCA-AMEA by starting from the polarized state in Fig.~\ref{fig:fig3}(c). (a) Nonequilibrium steady-state current as function of applied bias $\Phi$. (b) Orbital occupations as function of the applied bias. (c) Orbital spectra for $\Phi = 2 \mathrm{eV}$. (d) Same as (c) for $\Phi = 3.5 \mathrm{eV}$. Here $U = 5.5\, \mathrm{eV}$, $J = 0.75\, \mathrm{eV}$ and $T = 0.025\ \mathrm{eV}$.}
\label{fig:fig4}
\end{figure}

Because the spectrum is gapped, the current is expected to remain negligible until the applied bias compensates the band gap $E_{\mathrm{G}}$. Once $\Phi \gtrsim E_{\mathrm{G}}$, charge carriers can be injected across the gap and the current rises rapidly. This threshold-like behavior is clearly visible in Fig.~\ref{fig:fig4}(a): $J$ is essentially zero for $\Phi \lesssim E_{\mathrm{G}} \approx 2\,\mathrm{eV}$ and then increases exponentially as $\Phi$ is raised beyond $E_{\mathrm{G}}$. The inset highlights the many orders of magnitude spanned by the current.

For $\Phi \geq 2 \, \mathrm{eV} \approx E_{\mathrm{G}}$ the bias induces a redistribution of charge among the correlated orbitals. Due to its larger bandwidth, the in-plane $xy$ orbital responds first: electrons are promoted from its LHB into the corresponding upper band, so that the initial transfer is predominantly intra-orbital. Upon further increasing $\Phi$, electrons from the LHBs of the $yz$ and $xz$ orbitals are also promoted into the upper band of the $xy$ orbital, yet the orbital occupations remain approximately constant over an extended bias range, see Fig.~\ref{fig:fig4}(b). Only at larger bias, $\Phi \geq 1.7 E_{\mathrm{G}}$, does a net depletion of the $xy$ orbital in favor of $yz$ and $xz$ become apparent, see again Fig.~\ref{fig:fig4}(b).

These trends are reflected in the orbital-resolved DOS. We illustrate representative nonequilibrium spectra at $\Phi = 2\,\mathrm{eV} \approx E_{\mathrm{G}}$ and $\Phi = 3.5\,\mathrm{eV} \approx 1.7 E_{\mathrm{G}}$, corresponding to the onset of conduction and to a regime where the current already follows its exponential increase. At $\Phi = 2\,\mathrm{eV}$, the chemical potentials are just sufficient to start depleting the LHB and populating the upper band of the $xy$ orbital alone as can be observed in Fig.~\ref{fig:fig4}(c). At $\Phi = 3.5\,\mathrm{eV}$, one of the chemical potentials lies well within the LHBs of all orbitals while the other is at the edge of the conduction band of orbitals $yz/xz$, and the resulting nonequilibrium transfer predominantly injects electrons into the upper band of the $xy$ orbital, see Fig.~\ref{fig:fig4}(d). For $\Phi > 3.5 \mathrm{eV}$ the charge redistribution occurs from the $xy$ to the $yz/xz$ orbitals (spectra not shown), as evidenced by the occupation trends shown in Fig.~\ref{fig:fig4}(b)~\footnote{We note that the AMEA impurity solver has a finite energy resolution, which can lead to small numerical deviations from an exact Fermi distribution already at equilibrium and thus to a slight offset in the filling from the nominal value of $1/6$. Under nonequilibrium conditions, where the Keldysh component of the hybridization function acquires a nontrivial frequency dependence due to the application of a finite bias $\Phi$, these deviations may become more pronounced. As a consequence, the total occupation (per spin) can lie marginally above its nominal value even for biases $\Phi$ well within the gap, i.e., in a regime where physical charge injection is expected to be negligible.}.

\section{Conclusions}\label{sec:conclusions}

In this work we presented an approximate scheme to treat nonequilibrium steady states of impurity models with an arbitrary number of correlated orbitals and, via DMFT self-consistency, explore its application to the case of bulk and layered SrVO$_3$.
This mixed-configuration approximation (MCA) treats inter-orbital correlations beyond mean-field, while dealing with intra-orbital interactions exactly at the local level. The approach allows to replace the solution of the multi-orbital problem with a chain of single-orbital ones.
In this paper, the single impurity problem is addressed via the auxiliary master equation approach (AMEA), which maps the physical impurity to a finite open quantum system with bath orbitals and a Markovian environment and allows for an accurate and efficient solution both in equilibrium and steady-state nonequilibrium conditions.

We first applied our MCA-AMEA method to the case of bulk SrVO$_3$ and its layered structure in equilibrium conditions starting from the reciprocal space Hamiltonians obtained by means of the density functional theory. The MCA-AMEA recovers the predicted metallic phase of bulk SrVO$_3$ up to moderate values of the interaction strength, while overestimating the spectral weight of the lower band, thus favoring a metal-to-insulator phase transition as the interaction strength is further increased. On the other hand, all the results that can be found in the literature, among which we mention those obtained with quantum Monte Carlo (QMC) and fork tensor product states impurity solvers, predict the metallic phase to be stable up $U=5.5 \, \mathrm{eV}$ and beyond. Due to the fact that our approach works best for nondegenerate orbitals, this shortcoming is not surprising and additional work must be done to improve on this aspect.

We then applied our method to the case of a layered SrVO$_3$ in equilibrium. In this setup, the confinement of the $yz,xz$ orbitals to the $xy$ plane leads to the emergence of a crystal field that partly lifts the orbital degeneracy of the $t_{2g}$ manifold. The in-plane $xy$ orbital is then lowered in energy and retains the most part of the charge, giving rise to a charge-polarized state leading to the insulating phase. The results obtained within the dynamical mean-field theory (DMFT) using our MCA-AMEA scheme recover the charge polarization of the system in favor of the in-plane $xy$ orbital, albeit not as pronounced when compared to the prediction obtained by solving the DMFT self-consistent cycle with QMC. This fact suggests that our MCA approach needs to be refined in such a way to deal with (almost) degenerate orbital systems. 

One possible extension of the method is to formulate the MCA at the level of individual spin-orbital degrees of freedom by means of a cumulant expansion, following Refs.~\cite{kubo.62,sh.we.80}. In this way, one would explicitly treat connected clusters of spin and orbital flavors across different orbitals, including in particular smaller building blocks such as a fixed orbital and fixed spin flavor, as well as larger ones involving several flavors. The main difficulty, however, is that this construction tends to generate non-causal Green's functions, and hence non-causal spectra, already in the simplified case where spin-flip and pair-hopping terms are neglected. Work along these lines is in progress.

The solution of the one-shot impurity problem obtained with MCA-AMEA starting from the converged solution of the DMFT with QMC is in better agreement with the QMC results. Both the spectral features and the orbital densities show a stronger tendency towards charge polarization. This suggests that other factors such as a drift of the self-consistent solution obtained within DMFT might be at play. At the same time, for non-degenerate orbitals MCA-AMEA can provide an accurate and inexpensive one-shot impurity solution when the goal is to assess spectra and orbital occupations without performing a full DMFT self-consistency cycle.

Finally, when applied to a prototypical nonequilibrium setup, MCA-AMEA shows charge redistribution among the orbitals as a voltage bias is applied and correctly predicts a threshold-like behavior of the steady-state current as function of the bias. This shows that, even suffering from the limitations mentioned above, our method can be used to address nonequilibrium many-orbital lattice problems in a relatively computational cheap way. 
 

\begin{acknowledgments}
This research was funded by the Austrian Science Fund (FWF) [Grant DOI:10.55776/P33165], and by NaWi Graz. For the purpose of open access, the author has applied a CC BY public copyright licence to any Author Accepted Manuscript version arising from this submission. Results have been obtained using the A-Cluster at TU Graz as well as the Austrian Scientific Computing (ASC) infrastructure.
\end{acknowledgments}

\section*{Data Availability}
The data that support the findings of this study are available in the TU Graz Repository at \url{https://doi.org/10.3217/snf77-2hv59}.

\bibliography{references_database,my_refs}

\end{document}